\documentclass[preprintnumbers, floatfix, preprintnumbers, letterpaper, superscriptaddress,nofootinbib, twocolumn]{revtex4}
\pdfoutput=1
\usepackage{graphicx}
\usepackage{microtype}
\usepackage{amsmath}
\usepackage{amssymb}
\usepackage{subfigure}
\usepackage{hyperref}
\usepackage{url}
\usepackage{xcolor}
\usepackage{color}
\usepackage{mathrsfs}
\usepackage{calrsfs}
\usepackage{amsfonts}
\usepackage{latexsym}
\usepackage{ragged2e}
\usepackage{epsfig}
\usepackage{textcomp}

\usepackage{caption}
\DeclareCaptionJustification{justified}{\leftskip=0pt \rightskip=0pt \parfillskip=0pt plus 1fil}
\definecolor{vividviolet}{rgb}{0.62, 0.0, 1.0}
\definecolor{amaranth}{rgb}{0.9, 0.17, 0.31}
\definecolor{palatinateblue}{rgb}{0.15, 0.23, 0.89}
\definecolor{brightpink}{rgb}{1.0, 0.0, 0.5}
\definecolor{cornflowerblue}{rgb}{0.39, 0.58, 0.93}
\definecolor{deepcarminepink}{rgb}{0.94, 0.19, 0.22}
\definecolor{radicalred}{rgb}{1.0, 0.21, 0.37}

\hypersetup{ linktoc=all,
    colorlinks, linkcolor={palatinateblue},
    citecolor={brightpink}, urlcolor={amaranth}
}

\graphicspath{{Images/}}

\makeatletter
\def\@fnsymbol#1{{\ifcase#1\or \textleaf  \else\@ctrerr\fi}}
	\makeatother

\begin{document}

\title{GUP-Corrected Black Hole Thermodynamics and the Maximum Force Conjecture}

\author{Yen Chin \surname{Ong}}
\email{ycong@yzu.edu.cn}
\affiliation{Center for Gravitation and Cosmology, College of Physical Science and Technology,\\ Yangzhou University, Yangzhou 225009, China}
\affiliation{School of Physics and Astronomy,
Shanghai Jiao Tong University, Shanghai 200240, China}
\affiliation{\footnotesize {Nordita, KTH Royal Institute of Technology \& Stockholm University,
Roslagstullsbacken 23, SE-106 91 Stockholm, Sweden}}

\begin{abstract}
We show that thermodynamics for an asymptotically flat Schwarzschild black hole leads to a force of magnitude $c^4/(2G)$. This remains true if one considers the simplest form of correction due to the generalized uncertainty principle. We comment on the maximum force conjecture, the subtleties involved, as well as the discrepancies with previous results in the literature.
\end{abstract}

\maketitle

\section{Black Hole Thermodynamics and Maximal Force Conjecture}

Consider an asymptotically flat Schwarzschild black hole. The first law of black hole thermodynamics is $\text{d}M=T \text{d}S$. This allows us to define a force \cite{1504.01547} (see also \cite{1002.4278} in the context of cosmology)
\begin{equation}\label{F1}
F=T\frac{\text{d}S}{\text{d}r_h}=\frac{\text{d}M}{\text{d}r_h}=\frac{c^4}{2G},
\end{equation}
where $r_h$ denotes the horizon, and $T$ and $S$ are, respectively, the Hawking temperature and the Bekenstein-Hawking entropy, given explicitly by
\begin{equation}
T = \frac{\hbar c^3}{8\pi k_B GM}, ~~S=\frac{k_B c^3 A}{4\hbar G}.
\end{equation}
In this work we will refer to $F$ as the ``thermodynamics force'', since it has the correct dimension ${c^4}/G$ for a force. 

This quantity is reminiscent of -- and closely related to --  the ``entropic force'', proposed by Verlinde \cite{1001.0785}, but there are some subtle differences. Firstly, the entropic force describes the force on a test mass $m$ on a ``holographic screen'' at a distance $\Delta r$ from the gravitating body, which induces a change on the entropy of the holographic screen by $\Delta S = 2\pi k_B mc \Delta r/\hbar$. The entropic force, $F_\text{ent}$, satisfies
\begin{equation}\label{entropic}
F_\text{ent}\Delta r = T \Delta S,
\end{equation}
or equivalently,
\begin{equation}\label{entropic2}
F_\text{ent}\ = \frac{2\pi k_B mc T}{\hbar} .
\end{equation}
However the thermodynamics force defined above satisfies 
\begin{equation}
\frac{\text{d}S}{\text{d}r_h}=2\pi r_h = 4 \pi M,
\end{equation} 
which is independent of the mass $m$ of any test body\footnote{To put this in another perspective, consider the test mass to be a black hole of mass $m$, then $S_M=4\pi M^2$, $S_m=4\pi m^2$. Consider dropping the test mass into the black hole. Then $S_{M+m}=4\pi(M+m)^2$. The change in the entropy can be obtained from taking the limit:
\begin{equation}
\frac{\text{d}S}{\text{d}r_h} = \lim_{\Delta r \to 0} \frac{S_{M+m}-S_M}{2m}=2\pi r_h,\notag
\end{equation}
since taking $\Delta r \to 0$ is equivalent to taking $m \to 0$. That is, the ``thermodynamic force'' already considers small $m$ limit, whereas in what usually referred to as the ``entropic form'', $m$ is finite.
}. 
We also note that one could calculate a force from Eq.(\ref{entropic2}) by substituting in the Hawking temperature.
We obtain 
\begin{equation}\label{force3a}
\tilde{F} = \frac{2\pi k_B mc T}{\hbar} = \frac{c^4}{4G} \frac{m}{M}.
\end{equation}
The Hawking temperature is of course the temperature measured by an observer at infinity. For a local observer at finite $r$,
one should use the local (Tolman) temperature, $T_L$, and arrive at \cite{1002.0871}
\begin{equation}\label{force3b}
\tilde{F} = \frac{2\pi k_B mc T_L}{\hbar} = \frac{1}{\sqrt{1-\frac{r_h}{r}}}\frac{c^4}{4G} \frac{m}{M}.
\end{equation}
The series expansion of Eq.(\ref{force3b}) does not recover the Newtonian gravitational force law $F=GMm/r^2$, hence Myung \cite{1002.0871} argued that this expression (and so also Eq.(\ref{force3a})) has nothing to do with the entropic force (which provides a ``derivation'' of Newtonian gravity \cite{1001.0785}), thus we have denoted it with $\tilde{F}$ here, instead of $F_{\text{ent}}$. 

We emphasize that while the idea of gravity being an entropic force is still being debated (for numerous objections, see among others, \cite{1009.5414, 1108.4161, 1405.6909, 1601.07558}) and no consensus has been reached, Eq.(\ref{F1}) is just an established property of black holes, equivalent to the first law of black hole thermodynamics. In view of these subtleties, we do not refer to Eq.(\ref{F1}) as the entropic force, although we note that this quantity defined in the context of cosmology was referred to as such \cite{1002.4278}.

In general relativity, it has been conjectured that there exists an upper bound for forces acting between two bodies. In 4-dimensions, $F_\text{max}=c^4/{4G} \approx 3.026 \times 10^{43}$ N. Note that $c^4/G$ gives the correct physical dimension for a force, but the factor $1/4$ is nontrivial, though a variety of systems seem to agree with such a bound. See Gibbons \cite{Gibbons:2002iv}, Schiller \cite{chris}, and Barrow and Gibbons \cite{Barrow:2014cga}. This is known as the \emph{maximum force conjecture}, or maximum tension conjecture.  In \cite{Gibbons:2002iv}, it is further remarked that the factor $1/4$ might need to be revised subject to future research, but the main idea is that general relativity should have an upper bound for force, just like in special relativity there is a maximum allowed speed $c$. We therefore distinguish between two different forms of the maximum force conjecture:

\begin{itemize}
\item[(1)] \underline{Maximum Force Conjecture (Strong Form):}\newline In 4-dimensions, forces are bounded from above by $F_\text{max}=c^4/(4G)$.

\item[(2)] \underline{Maximum Force Conjecture (Weak Form):}\newline In 4-dimensions, there exists a positive number $K < \infty$, such that forces are bounded from above by $F_\text{max}=c^4K/G$. It is also possible that $K$ is a supremum instead of a maximum. 
\end{itemize}

That is to say, the weak form is simply the statement that forces cannot be arbitrarily large in general relativity, but makes no claim to the value of the lowest upper bound. The force defined in Eq.(\ref{force3a}), for example, satisfies the strong form of maximum force conjecture, since for any test mass $m \ll M$, we have $\tilde{F} \ll c^4/(4G)$. On the other hand the force defined in Eq.(\ref{F1}) only satisfies the weak form of the conjecture, though its deviation from the strong form is only of order unity (in Planck units).

Black hole physics is expected to receive quantum corrections as the black hole size reduces to near Planck length, as the result of losing mass via Hawking radiation. While a full quantum gravity theory is still lacking, there exist phenomenological models that allow us to study at least some properties of quantum black holes. The generalized uncertainty principle (GUP) is one such widely studied model. Since the thermodynamic force is related to the first law of black hole mechanics, it is interesting to explore how it is affected by GUP correction, which is known to modify both the Hawking temperature and the Bekenstein-Hawking entropy. This is the focus of this work.

\section{GUP-Corrected Black Hole and Maximal Force Conjecture}

From now on we will work with the Planck units in which $G=c=\hbar=1$. We also set $k_B=1$. As mentioned, in this work 
we would like to calculate the thermodynamic force as defined in Eq.(\ref{F1}), but in the context of GUP-corrected Schwarzschild black hole. In GUP the Heisenberg's uncertainty principle receives a correction term from gravitational effect:
\begin{equation}\label{GUP}
\Delta x \Delta p \geqslant \frac{1}{2} \left[\hbar + \frac{\alpha L_p^2 \Delta p^2}{\hbar}\right],
\end{equation}
where $\alpha$, the GUP parameter, is often taken to be $O(1)$ in theoretical considerations.
GUP is usually treated phenomenologically to model quantum gravitational effects \cite{9301067,9305163,9904025,9904026}. In the context of black hole physics, GUP correction yields \cite{pc}
\begin{equation}\label{TGUP}
T[\alpha]=\frac{M}{4\alpha \pi}\left(1-\sqrt{1-\frac{\alpha}{M^2}}\right)
\end{equation}
for the Hawking temperature. The negative sign in front of the square root was chosen so that the $\alpha \to 0$ limit recovers the standard Hawking's result \cite{pc}.

Applying the first law of black hole thermodynamics, one obtains the GUP-corrected Bekenstein-Hawking entropy
\begin{flalign}
S[\alpha]&=\int{\frac{1}{T}}\text{d}M \notag \\&= 2\pi \left[M^2 + M\sqrt{M^2-\alpha} - \alpha \ln(M+\sqrt{M^2-\alpha})\right] \notag \\&~~~+ \text{const.}
\end{flalign}
Note the appearance of the logarithmic term, which seems to be a common features in various quantum gravitational models (see, e.g., \cite{9701106, 0002040, 0111001, 0401070}). 
Note that in deriving Eq.(\ref{TGUP}), it is assumed that the uncertainty in the photon position is $\Delta x \sim r_h$, about the size of the black hole horizon. 
The horizon position is assumed to be still\footnote{Thus $S[\alpha]=A/4$ does \emph{not} hold.}  $r_h = 2M$. Thus we should expect that the thermodynamics force Eq.(\ref{F1}) is \emph{unchanged} since $\text{d}M/\text{d}r_h=1/2$.

This contradicts the claim in \cite{1504.01547}, in which it is claimed that the thermodynamics force (called ``entropic force'' therein) violates the (weak form -- and thus also strong form -- of) maximum force conjecture.
We should thus check the calculation explicitly for consistency\footnote{Note our $\alpha$ is the $\alpha^2$ of \cite{1504.01547}.}. Indeed,
\begin{equation}
\frac{\text{d}S[\alpha]}{\text{d}r_h}=\frac{2\pi (M^2+M\sqrt{M^2-\alpha}-\alpha)}{\sqrt{M^2-\alpha}},
\end{equation}
which upon multiplying with the GUP-corrected temperature Eq.(\ref{TGUP}), we obtain \emph{exactly} $1/2$, independent of the value (and sign) of  $\alpha$, consistent with our previous observation.

In terms of series expansion (asymptotic series in $M$), we have
\begin{flalign}\label{Texpand}
T[\alpha]&=\frac{1}{8\pi M} + \frac{1}{32}\frac{\alpha}{\pi M^3} + \frac{1}{64}\frac{\alpha^2}{\pi M^5} + \cdots \notag \\
&=T\left(1+\frac{\alpha}{4M^2}+\frac{\alpha^2}{8M^4} +\cdots \right) \notag \\
&=T\left(1+16\pi^2\alpha T^2 + 512\pi^4\alpha^2 T^4 + \cdots \right),
\end{flalign}
where $T=T[\alpha=0]$ is the original Hawking temperature $T=1/(8\pi M)$.
Similarly, the series expansion of the GUP-corrected entropy is
\begin{flalign}
S[\alpha] &= S\left(1-\frac{\alpha}{2M^2}\ln{M} + \frac{\alpha^2}{16 M^4} + \cdots \right) \notag\\
&=S\left(1-\frac{\pi\alpha}{S}\ln{S} + \frac{\alpha^2 \pi^2}{S^2} + \cdots\right),
\end{flalign}
where $S=S[\alpha=0]$ is the original Bekenstein-Hawking entropy $S=4\pi M^2$. 
Consequently,
\begin{equation}\label{FA}
F[\alpha]=F\left(1-\frac{\alpha\pi}{S} + 16 \pi^2 \alpha T^2 + \cdots \right),
\end{equation}
where $F=1/2$ is the original thermodynamic force defined in Eq.(\ref{F1}).
Although $F[\alpha]$ depends on $S$ and $T$, this series cannot possibly diverge (we know that it is exactly equal to $1/2$), as claimed in \cite{1504.01547}.
The apparent divergence comes from taking the limit $S \to 0$ and $T \to \infty$, which corresponds to the end stage of Hawking evaporation. 
With GUP correction both $T$ and $S$ are not allowed to tend to these limits. This can be appreciated from, e.g., Eq.(\ref{Texpand}): RHS consists of series of $T$, which would diverge if $T$ is allowed to diverge as in the usual scenario of Hawking evaporation. However, we know that GUP provides an upper bound for Hawking temperature (how this is achieved depends on whether $\alpha$ is positive or negative, see \cite{yco}). Therefore we must be careful about the range $M$ can take in the expansion, it should be the same as that of the defining equation itself.  In fact, the situation is better still: the terms
\begin{equation}
-\frac{\alpha\pi}{S} + 16 \pi^2 \alpha T^2
\end{equation}
in Eq.(\ref{FA}) \emph{cancel exactly}. The same holds for the higher order terms in the series expansion, so that $F \equiv F[\alpha], \forall \alpha$. 

We note that there are also typos in the series expansions in \cite{1504.01547}, which were likely transferred from the last reference therein, i.e., Tawfik et al. \cite{1502.04562} (Eq.(34) in the arXiv version). The mistake seems
to have been corrected in the next paper of partially the same authors of \cite{1504.01547}: A. Alonso-Serrano et al. \cite{1801.09660}. 
However, series expansion was again used in that work. It should again be emphasized that using
the \emph{full expression} for the Hawking temperature while defining the thermal wavelength would yield a quantitatively different result, compared to that obtained from just the first few terms of the series expansion [Eq.(26) of \cite{1801.09660}]: namely the latter led to Fig.2 therein, in which the black hole mass can go to zero, whereas we should have a nonzero remnant mass at which point evaporation stops. See \cite{1806.03691} for more discussions of this technical difference. 

We thus conclude that the thermodynamic force of an asymptotically flat Schwarzschild black hole is indeed $1/2$, as one would expect from the assumption that $r=2M$ independent of GUP-correction, and the definition of the thermodynamic force. 

One might take a somewhat different perspective. If one assumes that the GUP-corrected Hawking temperature can be derived via the usual Wick-rotation trick from the modified Schwarzschild geometry \cite{1407.0113}
\begin{equation}
\text{d}s^2=-f(r)\text{d}t^2+f(r)^{-1}\text{d}r^2+r^2\text{d}\Omega^2,
\end{equation}
where 
\begin{equation}
f(r)=1-\frac{2M}{r}+\frac{\varepsilon M^2}{r^2}, ~~ |\varepsilon| \ll 1,
\end{equation}
then the horizon is located at
\begin{equation}
r_h = 2M \left(\frac{1+\sqrt{1-\varepsilon}}{2}\right).
\end{equation}
In such a scenario the entropic force receives a small correction, but it remains finite. 
(The GUP parameter $\alpha$ is negative in this model \cite{1407.0113}, which has some virtues including preventing white dwarfs from getting arbitrarily large \cite{yco}.)

\section{Discussion}

In this work, we have re-examined a ``thermodynamic force'' -- not quite Verlinde's ``entropic force'' though related to  it -- that can be defined via the first law of black hole thermodynamics. Such a force is identically equal to $1/2$ in Planck units. This satisfies the weak form of the maximum force conjecture, but not its strong form that proposes $F_\text{max}=1/4$. Let us emphasize that the original maximum force conjecture only considers forces that act \emph{between two bodies}. It is not so clear what these ``two bodies'' refer to in the context of black hole thermodynamics, so perhaps it is not too surprising that it does not satisfy the strong form of the conjecture, whereas the force defined by considering a test mass in Eq.(\ref{force3a}) does -- when the temperature is measured at spatial infinity. On the other hand, in the context of asymptotically flat Kerr black holes, we can define an effective spring constant \cite{1412.5432}: $k:=M\Omega_+^2$, where $M$ is the mass of the black hole, and $\Omega_+$ the angular velocity of the event horizon. 
In the extremal limit, it turns out that Hooke's law $F=kx$ gives exactly $1/4$, which saturates the strong form of the maximum force conjecture \cite{1412.5432}, although such a ``force'' does not seem to be acting between any two physical bodies either; it is just a convenient effective description. This ``harmonic force'' vanishes for a Schwarzschild black hole (since $k=0$), which is distinct from the thermodynamic force $F=1/2$. Does \emph{any} force naturally defined in the context of general relativity satisfies the maximum force conjecture? (One can of course multiply a force by any factor, hence the requirement that it should be ``naturally'' defined from the context is crucial.)

If one applies the generalized uncertainty principle to an asymptotically flat Schwarzschild black hole, we show that its thermodynamic force is still finite. We explained the error made in \cite{1504.01547}, which claimed that GUP-corrected black hole could violate even the weak form of the maximum force conjecture (i.e., forces can diverge). However, note that \cite{1504.01547} also discussed other gravity theories, and even considered the possibilities that the values of $c$ and $G$ may not be a constant. We do not entertain such possibilities of running $c$ and $G$ in this work.

It is possible to consider the entropic force \`{a} la Verlinde, to derive the correction to the \emph{Newtonian} force law between two bodies with GUP correction. This yields with $\alpha=1$  \cite{hsuan}, 
\begin{flalign}
F_\text{N}[\alpha=1]=&F_\text{N}\left\{1+\beta (2-\ln{\beta})   \right.\notag \\
&+\left.\beta^2[4-5\ln{\beta} +(\ln{\beta})^2] + \cdots \right\},
\end{flalign}
where $F_\text{N}=GMm/R^2$ and $\beta := G\hbar/(c^3R^2)$. Newtonian forces are of course not required to be bounded from above. 

Let us note that the GUP modifications of the entropy and temperature of black hole rely on taking the
microcanonical corrections, as opposed to the canonical one. This subtle point was recently raised in \cite{1801.09660}. What this statement means is that we consider quantum corrections to microstate counting, while keeping the horizon area fixed. This contrasts with the canonical correction, which considers the thermal fluctuation of the horizon area, not related to the fundamental degrees of freedom. Following \cite{1801.09660}, since the object of study is correction to black hole physics which becomes important near the Planck scale, it is arguably the more fundamental microcanonical correction that should be considered. However, it would be interested to see in future works how the canonical correction affects the maximal force conjecture. 

Lastly, although the current work concerns asymptotically flat Schwarzschild black holes, it is of some interest to comment on cosmology. An immediate observation is that Eq.(\ref{FA}) can presumably be applied to the Hubble horizon $r_H=c/H$, where $H$ is the Hubble parameter. This gives a non-zero GUP correction to the thermodynamic force in the context of cosmology. Secondly, there exist a vast literature of applying Verlinde's entropic force to cosmology. In such applications, the Hawking temperature of the associated cosmological horizon comes with a prefactor, usually denoted $\gamma$, which is related to the position of the holographic screen, see \cite{1003.4526, 1208.2482} and the references therein. The value of $\gamma$ is often taken to be of order unity in theoretical considerations, however in \cite{1503.08722}, observational fitting suggested that $\gamma$ is two to four magnitude smaller. This is expected to affect the associated force in Eq.(\ref{force3a}). Despite a similar prefactor being included in \cite{1504.01547} while discussing GUP-corrected black hole, such a factor should not be included in the context of \emph{thermodynamical force}, in which the event horizon is the object of study, not the holographic screen as in the case of entropic force. The distinction between the two forces is subtle yet important.

\begin{acknowledgments}
YCO thanks the National Natural Science Foundation of China (grant No.11705162) and the Natural Science Foundation of Jiangsu Province (No.BK20170479) for funding support.
He wishes to thank Brett McInnes for useful suggestions, and Michael Good for some musings on various types of  ``factor of 2'' problems in theoretical physics. YCO also thanks members of Center for Gravitation and Cosmology (CGC) of Yangzhou University (\href{http://www.cgc-yzu.cn}{http://www.cgc-yzu.cn}) for discussions. YCO also thanks Nordita, where this work was finalized, for hospitality during his summer visit.
\end{acknowledgments}

\end{document}